\documentclass{ws-procs975x65}

\def\beq{\begin{equation}}
\def\eeq{\end{equation}}

\begin{document}

\title{Constraining cosmic isotropy with type Ia supernovae}
\author{C. A. P. Bengaly Jr.\footnote{e-mail: carlosap@on.br}, A. Bernui\footnote{e-mail: bernui@on.br} and J. S. Alcaniz \footnote{e-mail: alcaniz@on.br}}
\address{Observat\'orio Nacional, 20921-400, Rio de Janeiro - RJ, Brasil\\
}

\begin{abstract}
We investigate the validity of the Cosmological Principle by constraining the cosmological parameters $H_0$ and $q_0$ through the celestial sphere. Our analyses are performed in a low-redshift regime in order to follow a model independent approach, using both Union2.1 and JLA Type Ia Supernovae (SNe) compilations. We find that the preferred direction of the $H_0$ parameter in the sky is consistent with the bulk flow motion of our local Universe in the Union2.1 case, while the $q_0$ directional analysis seem to be anti-correlated with the $H_0$ for both data sets. Furthermore, we test the consistency of these results with Monte Carlo (MC) realisations, finding that the anisotropy on both parameters are significant within $2-3\sigma$ confidence level, albeit we find a significant correlation between the $H_0$ and $q_0$ mapping with the angular distribution of SNe from the JLA compilation. Therefore, we conclude that the detected anisotropies are either of local origin, or induced by the non-uniform celestial coverage of the SNe data set. 
\end{abstract}


\bodymatter

\section{Introduction}

The Cosmological Principle (CP) is one the most fundamental hypothesis upon which the concordance model based. In this work, we discuss the validity of the cosmological isotropy with different compilations of Type Ia Supernovae (SNe), namely the Union2.1~\citep{union12}) and JLA data sets~\citep{jla14}, using a hemispherical comparison method, hence determining whether the cosmological isotropy actually holds in large angular scales, and whether such hypothesis is not only a mathematical simplification, but a valid assumption.

\section{Methodology}

\subsection{The Hubble-, q- and sigma-maps}

We test the isotropy of the Universe expansion by mapping the $H_0$ and $q_0$ parameters through the celestial sphere, so that an opposite hemisphere comparison is performed  following Ref.~\refcite{bernui08} (see also Ref.~\refcite{bernui12}). Each pair of these hemispheres is well defined by the HEALpix pixelization scheme~\citep{gorski05}, such that we fit $H_0$ and $q_0$ by minimising the following quantity

\begin{equation}
\label{eq:chi2}
\chi^2 = \sum_i\left(\frac{\mu_i-\mu_{\mathrm{th}}(z_i,\mathbf{p})}{\sigma_{\mu_i}}\right)^2 \;,
\end{equation}

\noindent where the set $(z_i, \mu_i, \sigma_{\mu_i})$ contains the observational information of the SNe data, i.e., redshift, distance moduli and associated uncertainty of the {\it{i-th}} object, respectively, where $\mu_{\mathrm{th}}(z,\mathbf{p})$ is the distance modulus given by a specific cosmological model according to

\begin{equation}
\label{eq:mu_th}
\mu_{\mathrm{th}}(z,\mathbf{p}) =
5\log_{10}{ [ D_L(z,\mathbf{p}) ] } + 42.38 - 5\log_{10}(h) \;,
\end{equation}

\noindent where $h \equiv H_0/100$, $H_0 \equiv 100 \,\mbox{Km} / \mbox{s} / \mbox{Mpc}$, and $D_L(z,\mathbf{p})$ is the adimensional luminosity distance, whose arguments are the redshift $z$, in addition to the set of cosmological parameters $\mathbf{p}$ which describe the underlying cosmological model~\footnote{As we restrict our analyses to $z \leq 0.20$, $D_L(z,\mathbf{p})$ is given by a cosmographic expansion up to second order, where $\mathbf{p} = \{H_0,q_0\}$.}.

\begin{figure*}[!t]
\includegraphics[scale=0.23, angle = +90]{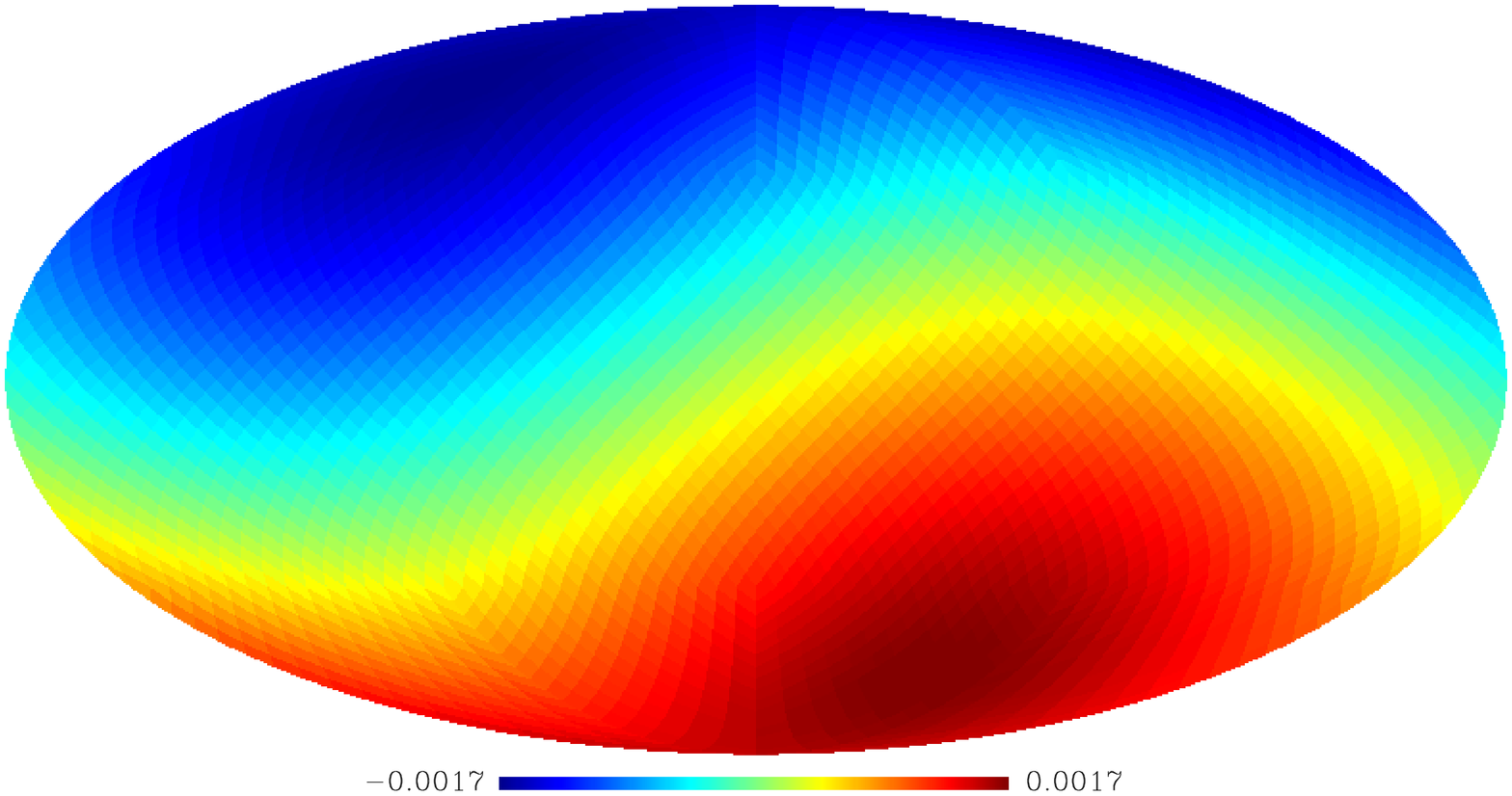}
\hspace{0.2cm}
\includegraphics[scale=0.23, angle = +90]{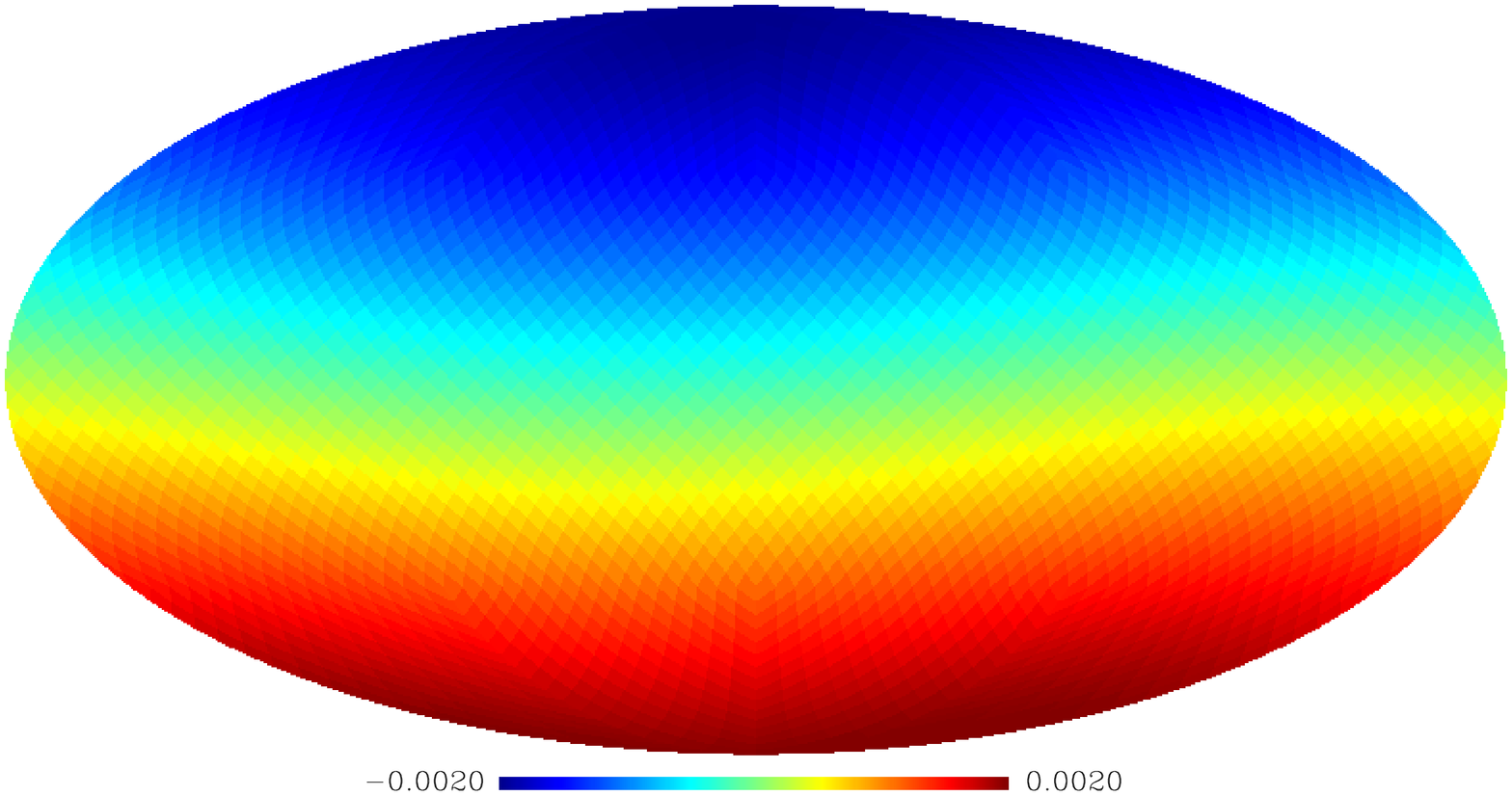}
\caption{{\it Left panel:} The result of the sigma-map analysis for the Union2.1 data set. {\it Right panel:} The sigma-map obtained for the JLA compilation. We show the dipole contribution of these maps, which shows that the maximal sigma-amp points towards $(l,b) = (295.00^{\circ}, \; -63.45^{\circ})$ (Union2.1) and $(l,b) = (195.00^{\circ}, \; -81.22^{\circ})$ (JLA).} 
\label{fig1}
\end{figure*}

\begin{figure*}[!t]
\includegraphics[scale=0.22, angle = -90]{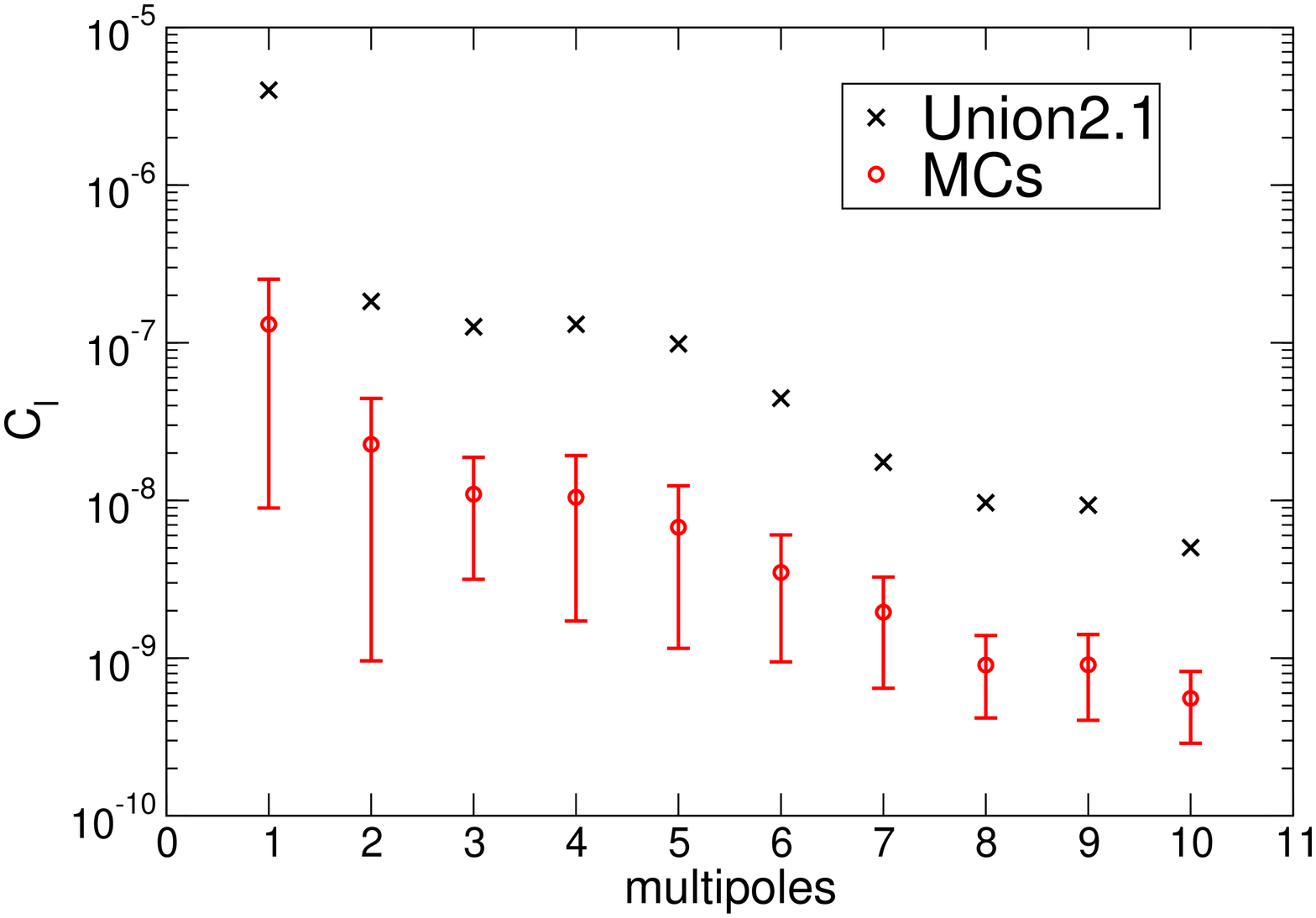}
\hspace{0.2cm}
\includegraphics[scale=0.22, angle = -90]{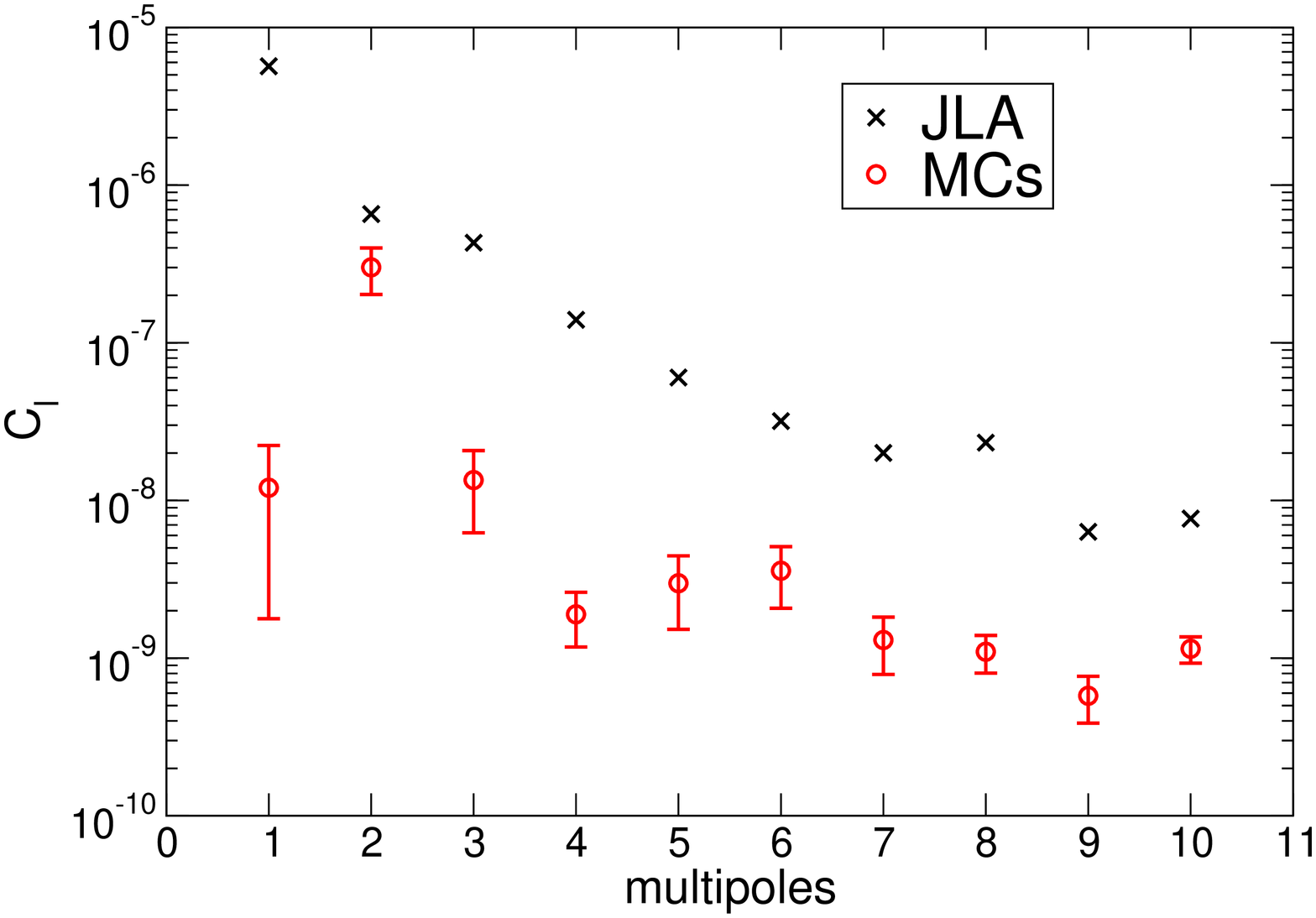}
\caption{{\it Left panel:} The angular power spectrum, $\{C_{\ell}\}$, up to $\ell = 10$, of the sigma-map for the Union2.1 catalogue. {\it Right panel:} the same for the JLA catalogue. The crosses represent the values for the original data sets, while the red circles assign the average spectra from sigma-maps obtained from 500 MC realisations. Their error bars are estimated using the median absolute deviation of each coefficient of these spectra.}
\label{fig2}
\end{figure*}

\begin{figure*}[!t]
\includegraphics[scale=0.23, angle = +90]{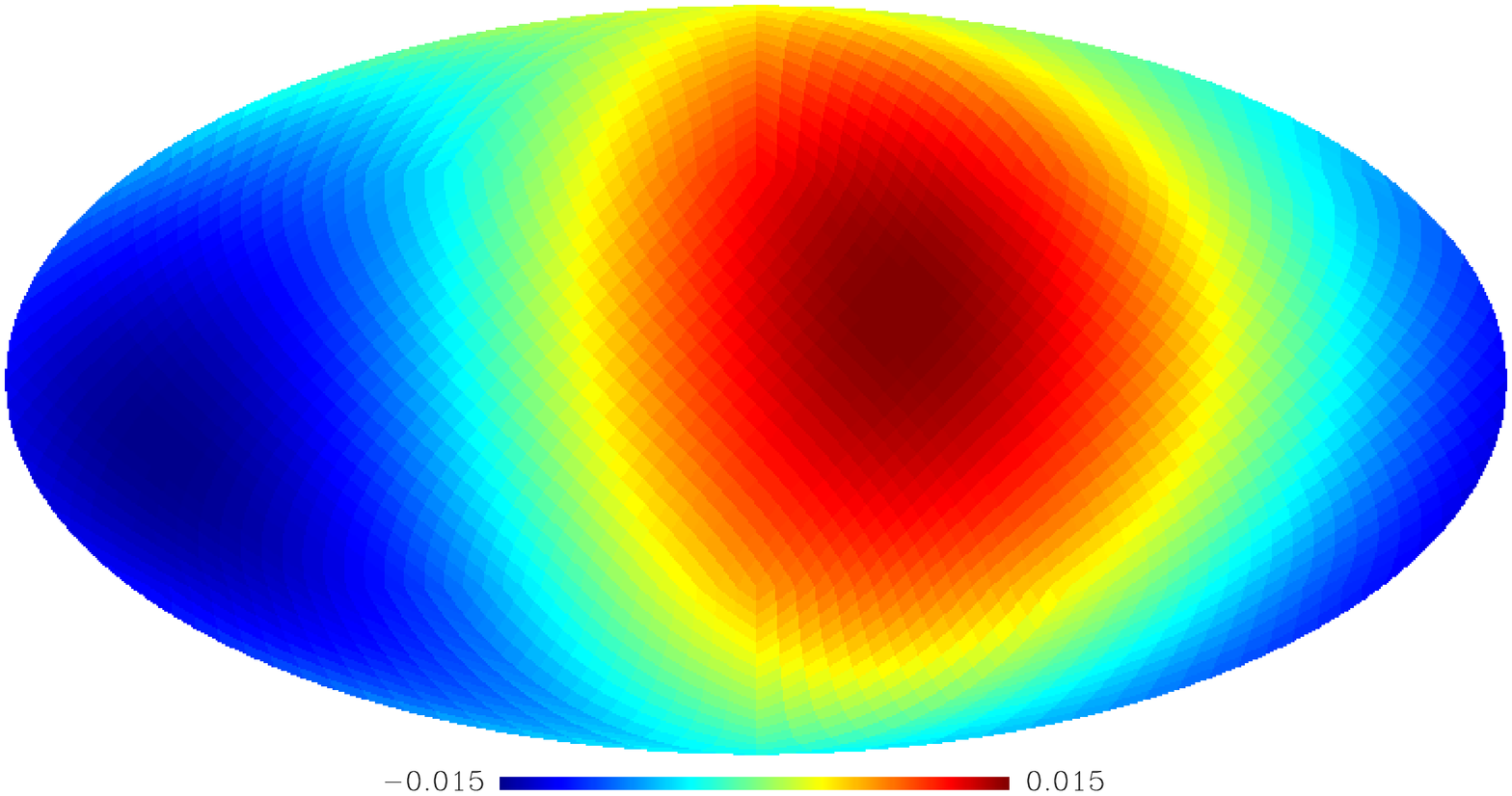}
\hspace{0.2cm}
\includegraphics[scale=0.23, angle = +90]{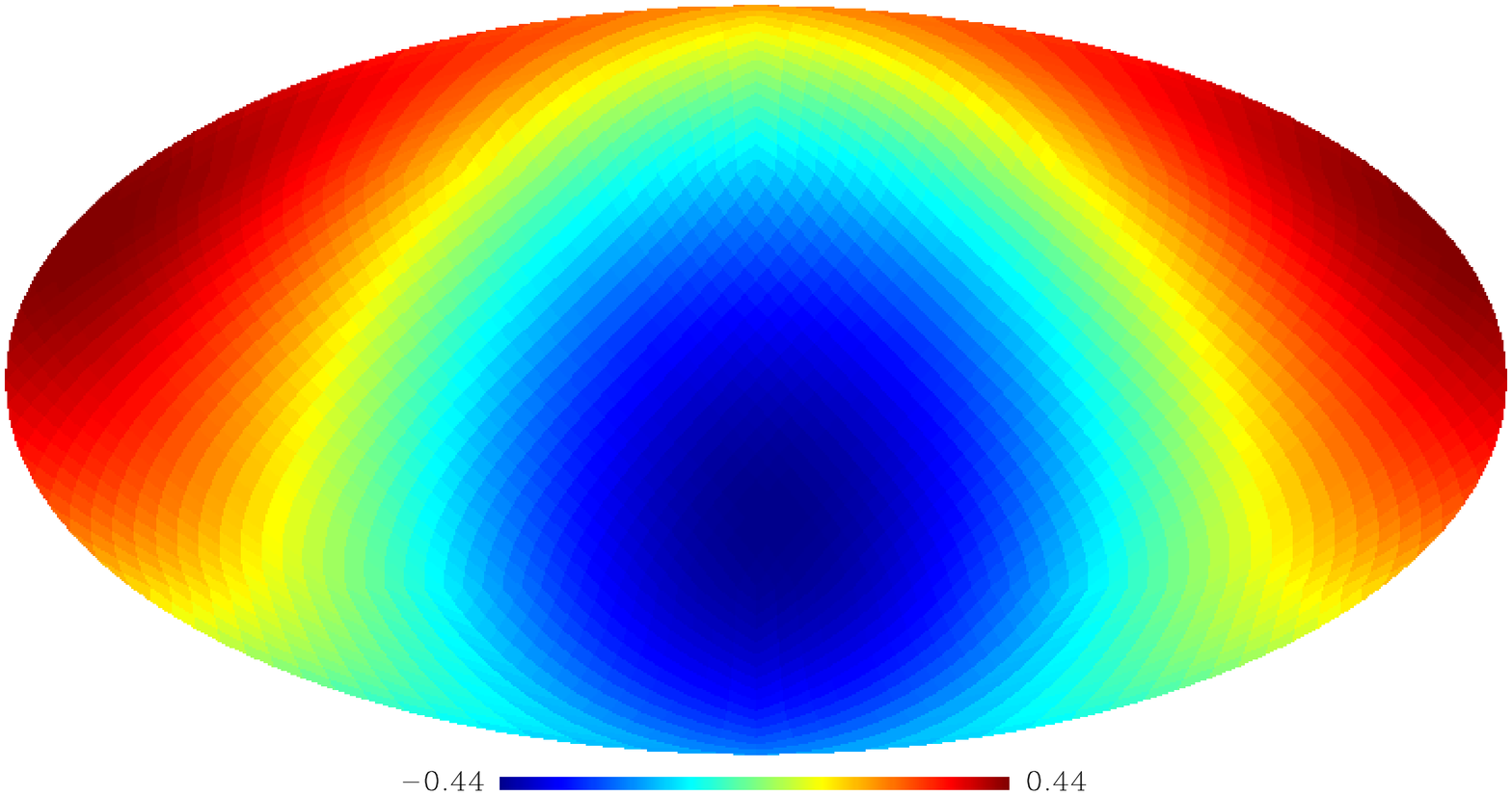}
\caption{{\it Left panel}: The dipole contribution of hubble-map analysis obtained from the Union2.1 catalogue. {\it Right panel}: The dipole contribution for the q-map analysis obtained from the same data set. The hubble-map maximal anisotropy is detected towards the direction $(l,b) = (326.25^{\circ}, \; 12.02^{\circ})$, while the q-map direction of maximal acceleration points at the $(l,b) = (354.38^{\circ}, \; -27.28^{\circ})$ direction.}
\label{fig3}
\end{figure*}

\begin{figure*}[!t]
\includegraphics[scale=0.23, angle = +90]{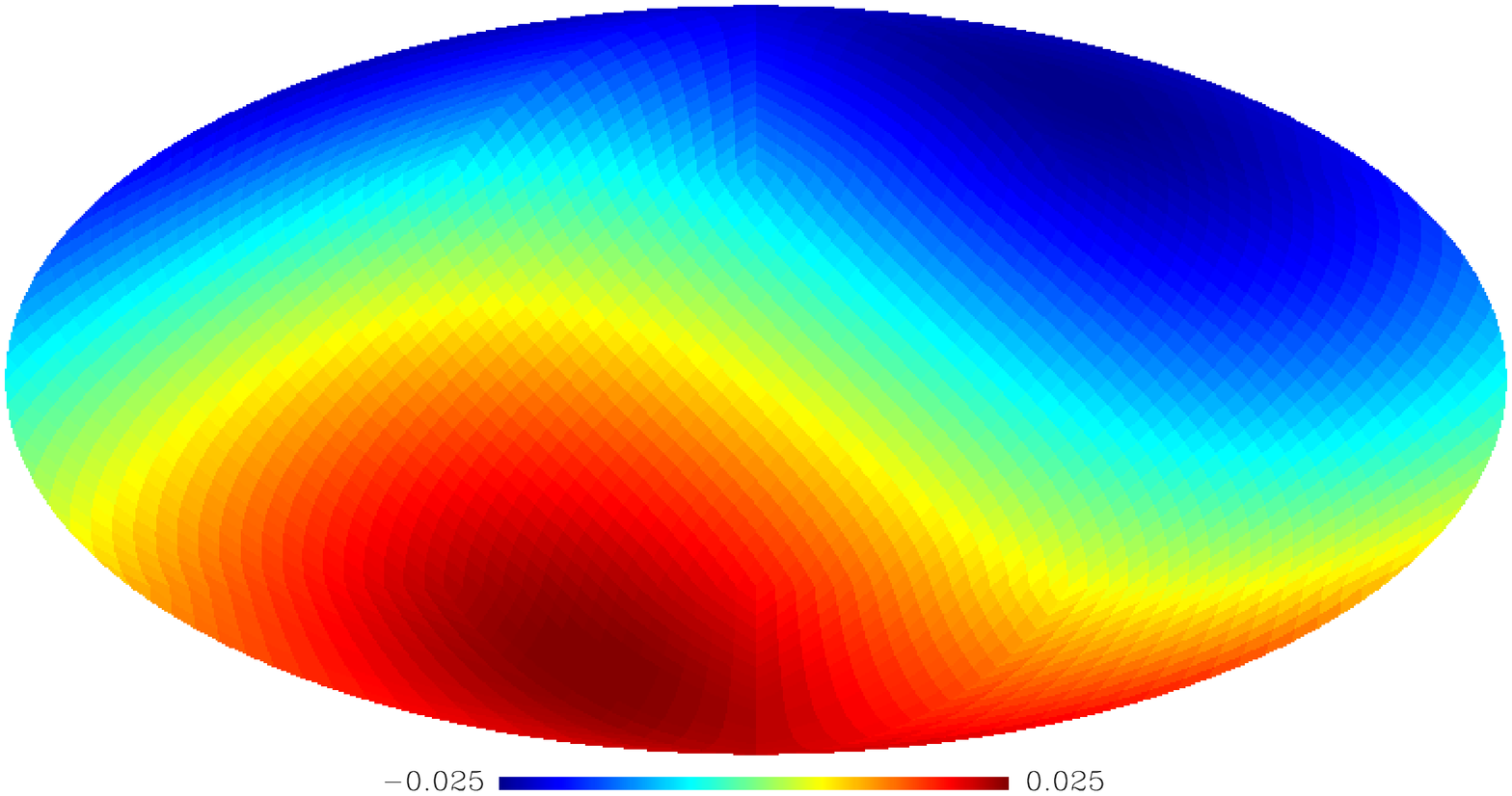}
\hspace{0.2cm}
\includegraphics[scale=0.23, angle = +90]{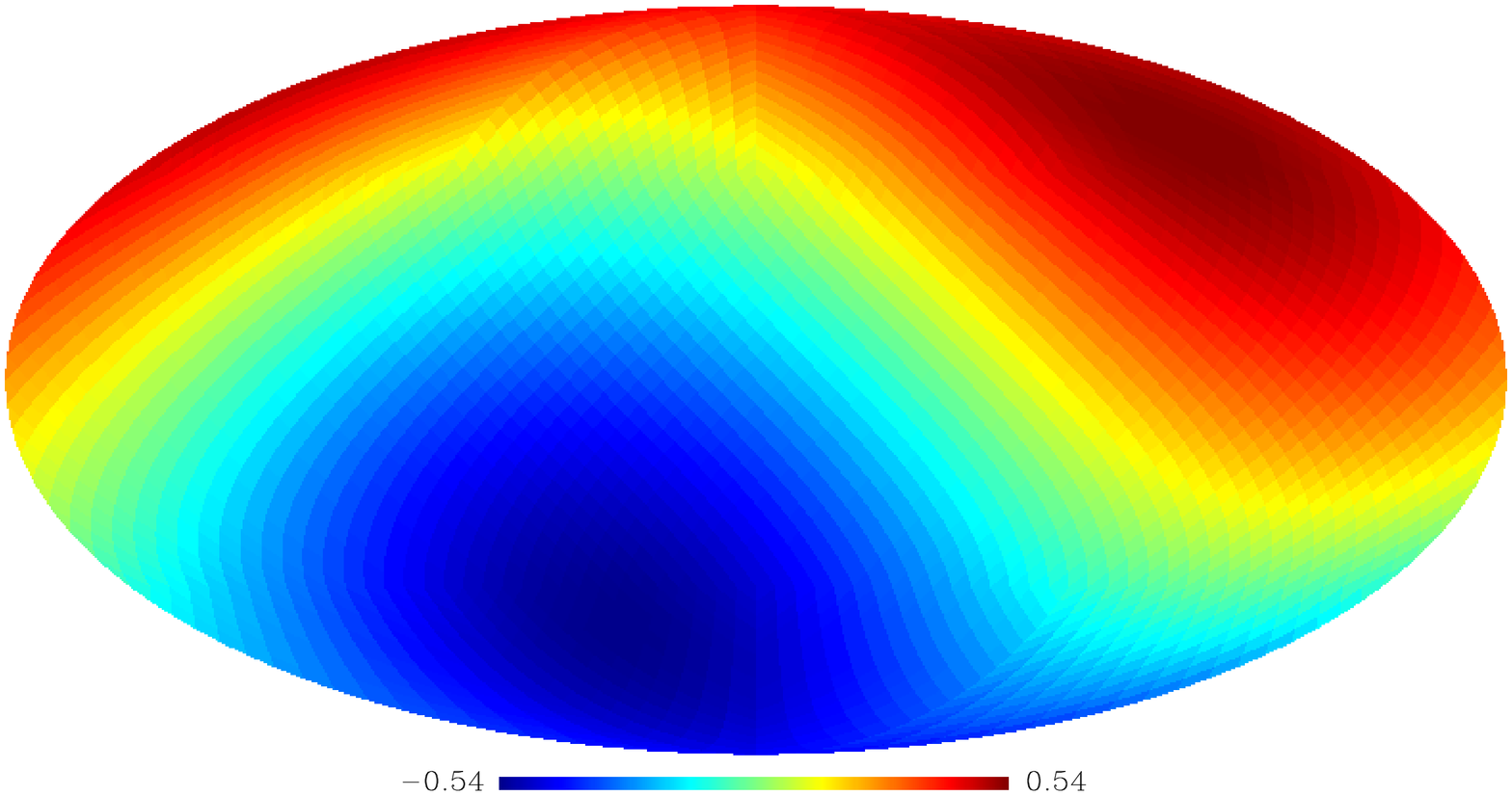}
\caption{{\it Left panel}: The same of the~\ref{fig3} left panel, but for the JLA case. {\it Right panel}: Also the same as~\ref{fig2}, but for the JLA data set The hubble-map maximal anisotropy is detected towards the direction $(l,b) = (58.00^{\circ}, \; -60.43^{\circ})$, while the q-map direction of maximal acceleration points at the $(l,b) = (225.00^{\circ}, \; 51.26^{\circ})$ direction.}
\label{fig4}
\end{figure*}

Furthermore, we quantify the angular non-uniformity of the data sets using the method named sigma-map, as performed in Ref.~\refcite{bernui08} (see also Ref.~\refcite{bengaly15a}), which is based upon the two-point angular correlation function of the cosmic objects distribution computed inside each assigned hemisphere. In other words, this estimator constructs a pixelised map in which its colour ranges from blue, when the actual distribution of SNe is less correlated than the mean value expected in a random catalogue, to red, in the case when the correlation is larger. In addition, we analyse the anisotropies of the cosmological parameters and the angular SNe distribution not only in the pixel space, but in the multipole space as well, so that $\delta(\theta,\phi) = \sum_{\ell,\, m} A_{\ell\, m} Y_{\ell\, m}(\theta,\phi)$~\footnote{$\delta(\theta,\phi)$ represents the quantity scanned through the celestial sphere, such as the $H_0$ and $q_0$ parameters.}, and $C_{\ell} \equiv (1 / (2\ell+1)) \sum_{m={\mbox{\small -}}\ell}^{\ell} \, |A_{\ell\, m}|^2$ is the angular 
power spectrum of the Hubble-, q- and sigma-map. Since we are interested in large scale angular correlations, we limit our analyses to $\ell \leq 10$. 

\subsection{Statistical significance tests}

The statistical significance of the Hubble and q-maps analyses is estimated with two different approaches. In the first approach, the galactic coordinates of each SNe is fixed, yet the set $(z, \mu, \sigma_{\mu})$ is shuffled (hereafter {\it shuffle} test). The second approach also keeps the original $(z, \mu, \sigma_{\mu})$ of each object yet the SNe positions are isotropically redistributed on the celestial sphere (hereafter the {\it MC} test). Hence, we can test whether the directional dependence of these parameters are statistically significant in its amplitude as well as in its direction. 

\section{Results}

\begin{figure*}[!t]
\includegraphics[scale=0.22, angle = -90]{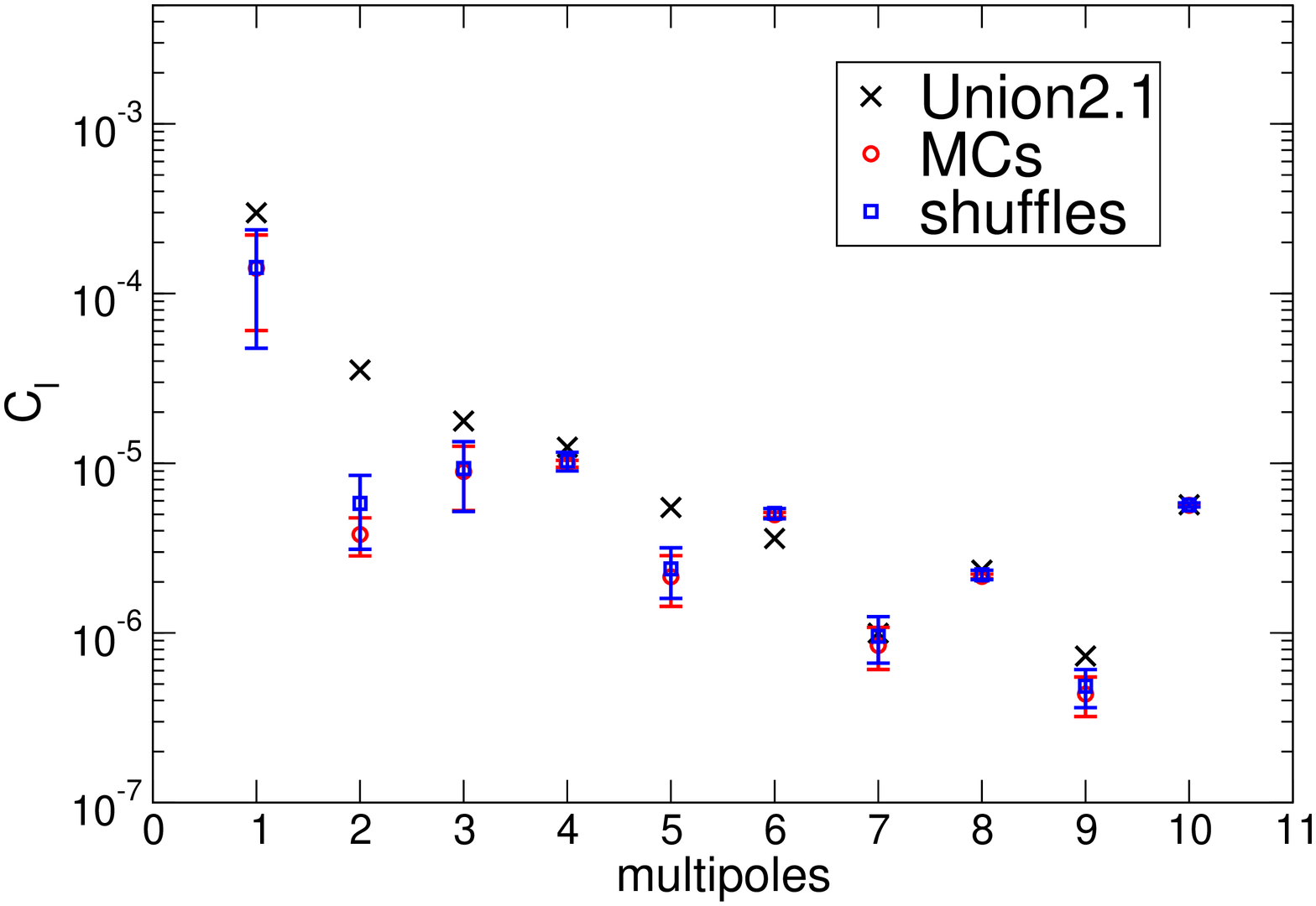}
\hspace{0.2cm}
\includegraphics[scale=0.22, angle = -90]{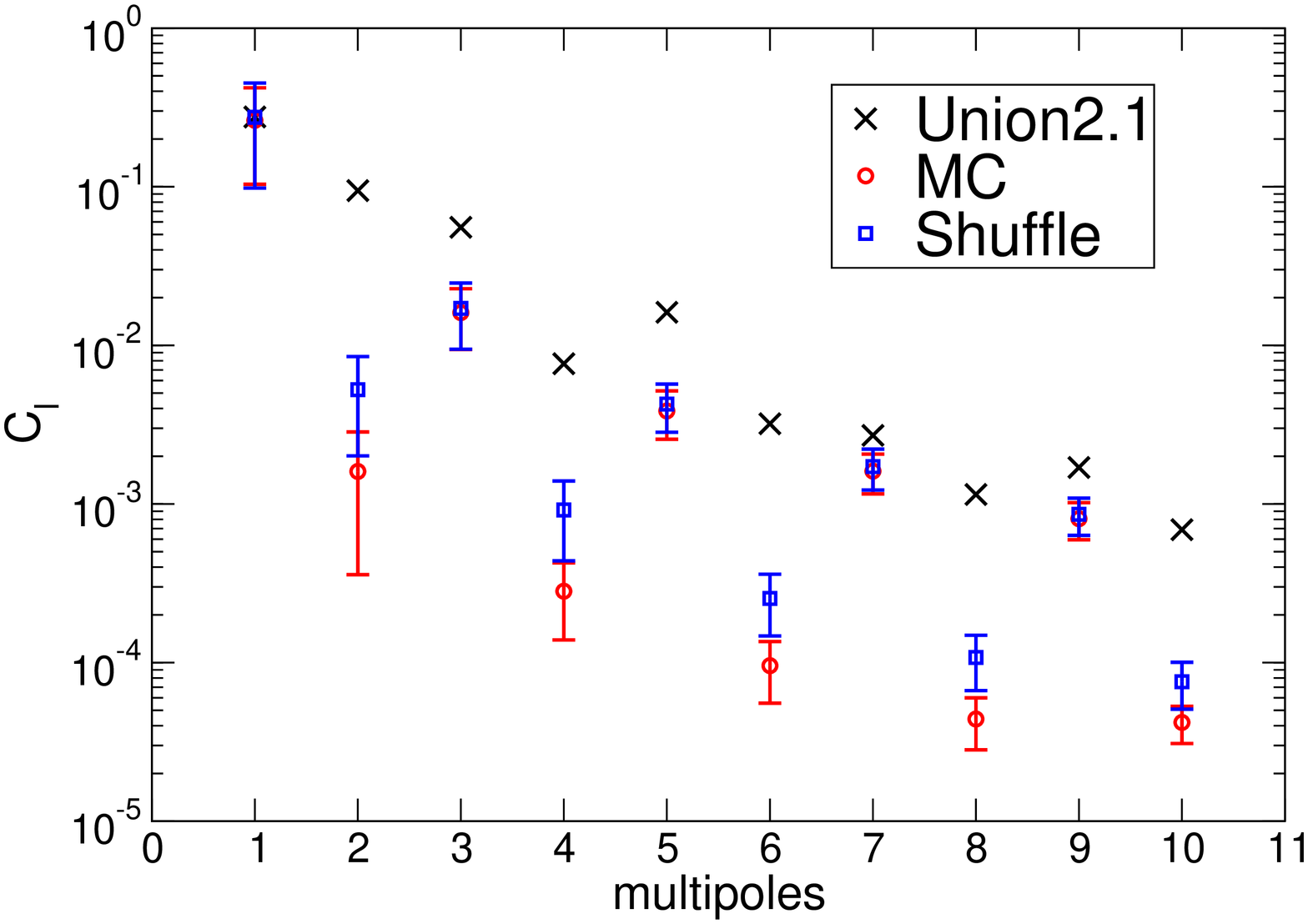}
\caption{{\it Left panel:} The angular power spectrum, $\{C_{\ell}\}$, up to $\ell = 10$,  of the Hubble-map for the Union2.1 catalogue. {\it Right panel:} the same for the q-map. 
The crosses represent the values for the original data set, while the red (blue) circles (squares) assign the average spectra from Hubble-maps obtained from 500 MC (shuffle) realisations. Their respective error bars are estimated using the median absolute deviation of each coefficient of these spectra.}
\label{fig5}
\end{figure*}

\begin{figure*}[!t]
\includegraphics[scale=0.22, angle = -90]{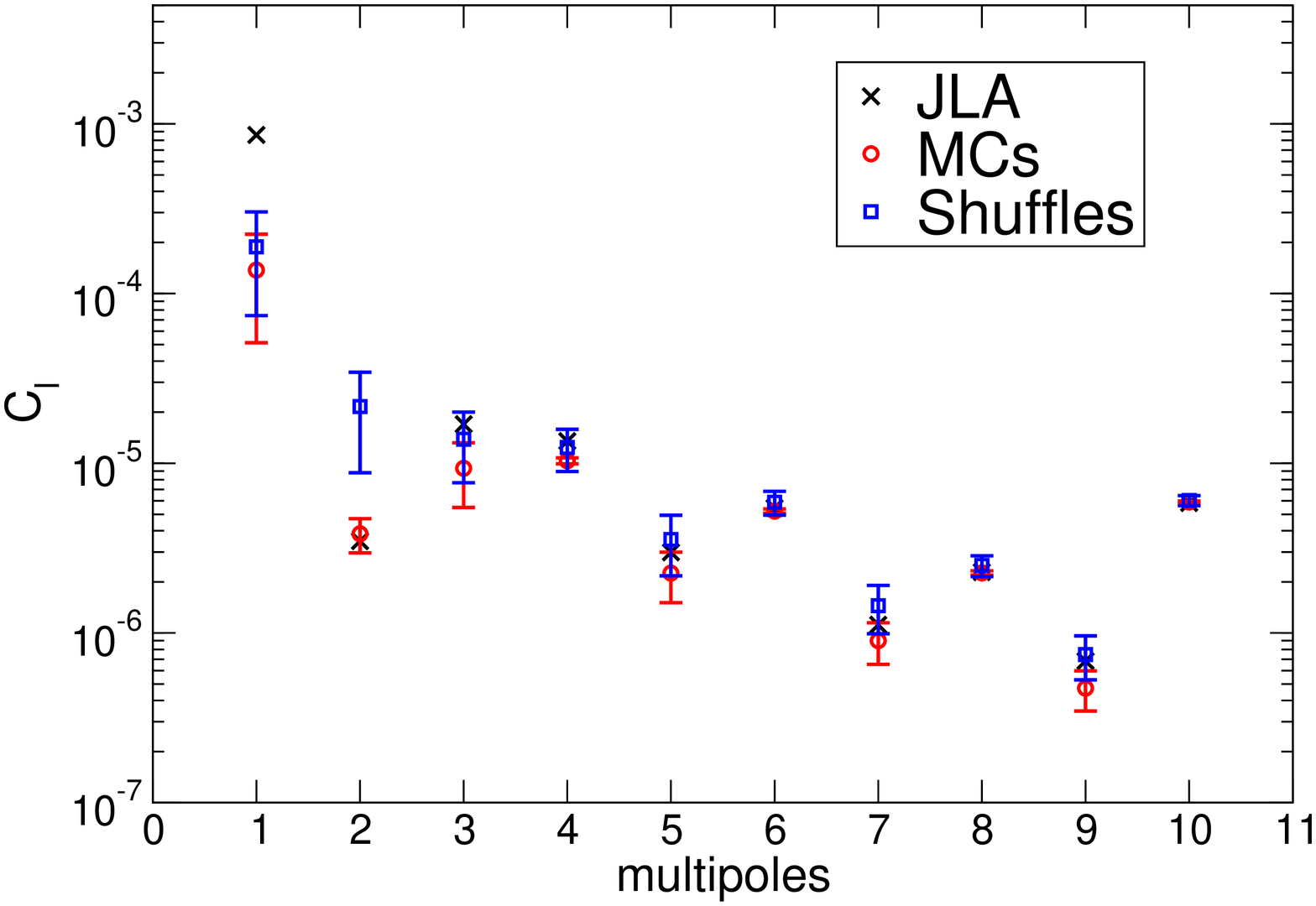}
\hspace{0.2cm}
\includegraphics[scale=0.22, angle = -90]{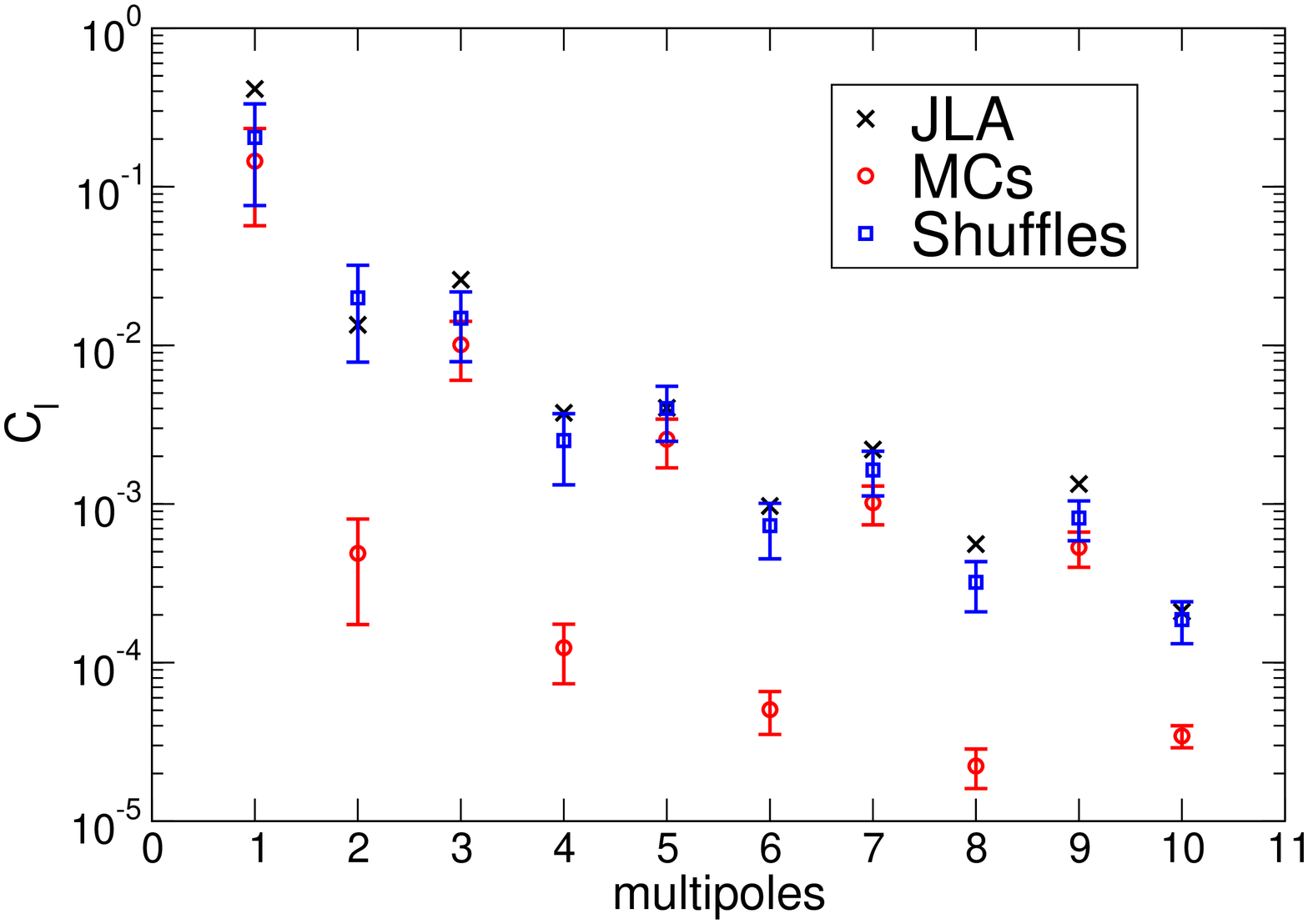}
\caption{{\it Left panel:} The angular power spectrum, $\{C_{\ell}\}$, up to $\ell = 10$, of the Hubble-map for the JLA catalogue. {\it Right panel:} the same for the q-map. 
The crosses represent the values for the original data set, while the red (blue) circles (squares) assign the average spectra from Hubble-maps obtained from 500 MC (shuffle) realisations. Their respective error bars are estimated using the median absolute deviation of each coefficient of these spectra.}
\label{fig6}
\end{figure*}

The results of the Sigma-map analyses are shown in Fig.~\ref{fig1} (pixel space) and Fig.~\ref{fig2} (multipole space) for both SNe data sets. It is possible to note that they present a preferred direction on the celestial sphere, as discussed on the description of Fig.~\ref{fig1}, and that the both SNe catalogues are highly inconsistent with a perfectly isotropic distribution, since the analyses performed in multipole space present much higher $\{C_{\ell}\}$'s than their average values obtained by the MCs. Moreover, the Hubble- and q-map results are featured in Fig.~\ref{fig3} (\ref{fig4}) for the Union2.1(JLA) compilations, for the analyses performed in pixel space, whereas Fig.~\ref{fig5} (\ref{fig6}) refer to the analyses carried out in multipole space for the Union2.1 (JLA) compilations as well. \\
We note that the direction $(l,b) = (326.25^{\circ}, \; 12.02^{\circ})$ obtained for the Union2.1 Hubble-map is consistent with the bulk flow motion direction estimated in Ref.~\refcite{turnbull12}, that is, $249 \pm 76$ km/s towards $(l,b) = (319^{\circ} \pm 18^{\circ}, 7^{\circ} \pm 14^{\circ})$, as well as many works which probed the cosmological isotropy with a similar approach~\citep{antoniou10, cai12, kalus13, chang14, lin15, bengaly15b, javanmardi15, carvalho15}. Moreover, the anisotropy of the $H_0$ can possibly explain the tension between the $H_0$ determinations~\citep{bengaly15c}) from low-$z$ standard candles~\citep{riess11} and Planck CMB temperature~\citep{planck15}. It was found that the maximal $H_0$ variance through the celestial sphere is consistent with their values, and that its direction is consistent with the bulk flow motion as well. This reinforces the idea that such anisotropy arises as a local effect, instead of an intrinsic cosmological anisotropy. We also evaluate the strength of the correlation between these maps, finding a negligible correlation between the Hubble- and q-maps with the sigma-map of the Union2.1 data set ($\rho = +0.059$ and $\rho = -0.200$, respectively), even though the correlation is moderate in the JLA analyses: $\rho=+0.651$ and $\rho=-0.446$, respectively. Thus, we conclude that the anisotropy detected on the Hubble and q-maps in the JLA data is possibly explained by the incompleteness of the sample in terms of sky coverage, while the anisotropy pointed by the Union2.1 SNe is most likely of local origin. \\

The results of the statistical significance are depicted, in multipole space, in Figs.~\ref{fig5} and~\ref{fig6} for the Union2.1 and JLA case, respectively. It is possible to note that the Union2.1 Hubble-map present mild disagreement with the {\it MC} and {\it shuffle} tests specially in the lower $\ell$ ($\ell < 5$), and that the q-map strongly disagrees with both realisations except for the dipole case, thus showing significant evidence for anisotropy in this analysis. Nevertheless, this signal can be ascribed to the limited constraining power of the Union2.1 data on this parameter, besides the degeneracy with the Hubble parameter. For the JLA catalogue, there is a better agreement between the real data and the {\it shuffle} realisations in both Hubble- and q-maps, except for the dipole contributions, whereas the data presents stronger disagreement with the {\it MC} runs specially in the q-map analysis. This result shows, once again, the fact that the angular non-uniformity of the JLA sample indeed biases 
the Hubble and q-map analyses.

\section{Summary}

We have shown that the anisotropy detected on the $H_0$ mapping with the SNe sample can be attributed to the bulk flow motion due to its proximity of reported directions in the literature, and the $q_0$ significant anisotropy probably arises due to the limitation of this data set. On the other hand, the JLA directional analyses show a significant dependence with its celestial coverage, then biasing the Hubble and q-map results. Therefore, we conclude that there is no significant violation of the cosmological isotropy with the latest SNe data in the $z \leq 0.2$ range, albeit next-generation surveys such as LSST and Euclid may improve this test with the greater precision and much larger data sets that they shall provide.

\section*{Acknowledgments}

We thank Roy Maartens, Ribamar Reis and Ivan Soares Ferreira for helpful discussions. We acknowledge the HEALPix package for the derivation of our analyses. This work is supported by  CNPq, FAPERJ and CAPES.

\end{document}